\documentclass[letterpaper]{article} 
\usepackage{aaai24}  
\usepackage{times}  
\usepackage{helvet}  
\usepackage{courier}  
\usepackage[hyphens]{url}  
\usepackage{graphicx} 
\usepackage{natbib}  
\usepackage{caption} 
\usepackage{algorithm}
\usepackage{algorithmic}

\usepackage{booktabs} 
\usepackage{subfigure}
\usepackage{multirow}
\usepackage{tabularx}
\usepackage{url}
\usepackage{hyphenat}
\usepackage{float}
\usepackage[table,xcdraw]{xcolor}
\usepackage{xspace}

\usepackage{newfloat}
\usepackage{listings}
\floatstyle{ruled}
\newfloat{listing}{tb}{lst}{}
\floatname{listing}{Listing}

\usepackage{paralist}

\usepackage{enumitem}

\usepackage{url}
\usepackage{graphicx}
\usepackage{color}
\usepackage{tablefootnote}
\usepackage{footnote}
\usepackage{tabularx}
\usepackage{textcomp}
\usepackage{caption}
\usepackage{subfigure}
\usepackage{footnote}
\usepackage{enumitem}
\usepackage{multirow}
\usepackage{amsmath}
\definecolor{Gray}{gray}{0.9}
\usepackage{colortbl}

\usepackage{newfloat}
\usepackage{listings}
\DeclareCaptionStyle{ruled}{labelfont=normalfont,labelsep=colon,strut=off} 
\floatstyle{ruled}
\newfloat{listing}{tb}{lst}{}
\floatname{listing}{Listing}
\usepackage{color}
\newif\ifcomment
\commenttrue
\ifcomment
\newcommand{\sz}[1]{{\bf \textcolor{red}{SZ: #1}}}
\else
\newcommand{\sz}[1]{}
\fi

\newif\ifcomment
\commenttrue
\ifcomment

\else
\newcommand{dn}[1]{}
\fi

\newif\ifrevision
\revisionfalse
\ifrevision

\newcommand{\revisioncomment}[1]{\textbf{\textcolor{red}{[#1]}}}
\else

\newcommand{\revisioncomment}[1]{}
\fi

\title{A Sticker is Worth a Thousand Words: Characterizing the Use\\ of Stickers in WhatsApp Political Groups in Brazil}

\author{
    Philipe Melo \textsuperscript{\rm 1, 2},
    João M. M. Couto \textsuperscript{\rm 1},
    Daniel Kansaon \textsuperscript{\rm 1},
    Vitor Mafra \textsuperscript{\rm 1}, \\
    Julio C. S. Reis \textsuperscript{\rm 2},
    Fabricio Benevenuto \textsuperscript{\rm 1}
}
\affiliations{
    \textsuperscript{\rm 1} Universidade Federal de Minas Gerais, Brazil\\
    \textsuperscript{\rm 2} Universidade Federal de Viçosa, Brazil\\

}

\usepackage{bibentry}

\begin{document}

\maketitle

\begin{abstract}
    With the increasing use of smartphones, instant messaging platforms turned into important communication tools. According to WhatsApp, more than 100 billion messages are sent each day on the app. Communication on these platforms has allowed individuals to express themselves in other types of media, rather than simple text, including audio, videos, images, and stickers. Particularly, stickers are a new multimedia format that emerged with messaging apps, promoting new forms of interactions among users, especially in the Brazilian context, transcending their role as a mere form of humor to become a key element in political strategy. In this regard, we investigate how stickers are being used, unveiling unique characteristics that these media bring to WhatsApp chats and the political use of this new media format. To achieve that, we collected a large sample of messages from WhatsApp public political discussion groups in Brazil and analyzed the sticker messages shared in this context\footnote{\textit{Warning!} This paper contains images and terms that may be potentially offensive.}.

\end{abstract}

\section{Introduction}

``\textit{A picture is worth a thousand words}'' is a popular expression that means that complex and sometimes multiple ideas can be transmitted by a single still image, which may convey its essence more effectively than a verbal description. Similarly, stickers are a form of visual media, popular in instant messaging services, through which users can express emotion, nonverbal language, and complex ideas \cite{wang2016more}. 
With the increasing importance of online communication in our daily lives, especially after the COVID-19 pandemic, users seek new and more efficient ways to express themselves through computers and smartphones, emphasizing the need for meaningful online emotional connections~\cite{liu2020relation}.
In this scenario, sometimes a simple text-based communication may not be enough to express every feeling we might want to convey. Even with emoji characters in text, this may prevent users from fully understanding the contextual subtlety and accuracy of the conversation. Users, however, may overcome such limitations with the usage of stickers~\cite{smileyface_stickers}, embodying some forms of interactions, gestures, and symbols with this emerging multimedia messaging format.

When it comes to WhatsApp, they introduced stickers on the platform referring to them as something to ``help you share your feelings in a way that you cannot always express with words''\footnote{\url{https://blog.whatsapp.com/introducing-stickers}}. WhatsApp allows users to create stickers of anything and for any occasion, and them to be quickly incorporated into various contexts on the platform, especially for political discussion, which is a popular topic on WhatsApp~\cite{Resende-WWW2019}. This becomes even more relevant considering that WhatsApp has recently gained attention due to campaigns that abuse this environment by quickly spreading misinformation on its network~\cite{melo2019complexnet} and playing a central role in the dissemination of fake news during political events~\cite{Bursztyn19thousand,resendewebsci19}. 

In Brazil, stickers had a significant increase in popularity, transcending their role as a mere form of humor to become a key element in political strategy. Throughout the 2022 presidential Brazilian election, stickers evolved into essential tools for disseminating political content across diverse groups with the goal of proliferating, capturing a broad audience, and circulating in groups that extended beyond the realm of fervent political activism \cite{figurinha_globo}. Such was their widespread, the Brazilian Electoral Court responded by creating a sticker pack, aiming to guide people in fact-checking news \cite{tse_sticker}. Moreover, since users are able to create their own stickers on WhatsApp, this permissive model is also being abused by users to create offensive and hateful stickers threatening other group members and creating an unsafe environment for the users. Recently, even the tool introduced by WhatsApp to automatically create stickers using AI was accused of generating xenophobic stickers within the app~\cite{ai_stickers_palestine}. 

Although stickers became very popular among users, we have observed this media being abused for a range of issues within WhatsApp, from misinformation campaigns to hateful harassing content. However, we still lack a systematic analysis of their usage on a large scale within the messaging platforms ecosystem due to the closed architecture of instant messaging platforms. 
In this study, we put forth a systematic characterization of sticker usage on WhatsApp, focusing specifically on their role within political public groups during the 2022 Brazilian elections. Our analysis aims to address the following research questions:
\begin{itemize}
    \item \textbf{RQ1:} Do users use stickers the same way they use other media formats on WhatsApp?; 
    \item \textbf{RQ2:} What are the characteristics of stickers shared on WhatsApp political groups in Brazil? 
    \item \textbf{RQ3:} How users are abusing of stickers to spread offensive content through WhatsApp?
\end{itemize}

To answer these questions, we collect a large-scale set of messages sent to a thousand political public groups on WhatsApp during five months of the Brazilian election period in 2022 (i.e., June to October). In total, we collected more than 6M messages, from which 650k are sticker messages. We also process distinct stickers, counting 57,031 unique media. 
In our initial efforts, we uncovered key insights indicating intriguing avenues for exploration in this content. Despite the greater prevalence of image and text messages, stickers are also extremely popular on WhatsApp. Our initial results suggest that stickers not only possess a distinct structure compared to other media formats but also reflect varied usage patterns among users. Furthermore, we discovered evidence of offensive and provocative stickers shared on public groups on WhatsApp, especially during the electoral period when public groups from opposing political parties suffered several attacks from users who invaded the group and spread hate speech.

\section{Background and Related Work} \label{sec:related}

The use of mobile instant messaging applications became popular has been changing the way users communicate online.
Compared to the traditional short messaging service (SMS), instant messaging offers a more social, conversational, and informal communication environment~\cite{whatsupwhatsapp13mobilechi}. Therefore, smartphone users felt closer to their correspondents through the exchange of messages~\cite{emojis_chi}.
This change in the way people connect also came with more complex ways of expressing, other than emoticons and emojis, influencing the emergency of stickers.

Stickers are usually described as emoticons in the form of colored images~\cite{chen21multimedialanguage}, as both can be used as visual reaction messages or determine the feeling of the interlocutor during the conversation, but differ from in-line emoticons and emojis in diversity, complexity, and usage~\cite{smileyface_stickers}, providing a richer communication~\cite{wang2016more}. Also, they differ from simple static images as they can be short animated movies, similar to a GIF.
Stickers are also closely related to memes~\cite{wang2019culturally}, which often present a combination of a picture and a statement, typically with sarcastic or humorous intention~\cite{davison2012language}. Usually, there is a template-based image that people modify, edit, and publish their version while keeping some key aspects so that the meme template is still recognizable.
Both memes and stickers can be created based on personal experiences, but inspiration can be obtained from popular cultural products such as television shows and video games~\cite{ge2020memeticstickers}. Additionally, users publish memes on the Web to create an incentive for others to share by replicating the original~\cite{mememagic2021cscw}.

The origin of these stickers dates back to 2011 with LINE in 2011, a Korean instant massaging application that got popular in Japan, mixing cartoons and ``emojis''\cite{stickers_origins}.
This tool allowed users to express social intention, identities, context, and social-cultural differences in a more specific way, inspiring other applications to adopt stickers as well. In 2013, Facebook added stickers to its platform, bringing the idea to the western market, but still with a limited set of precreated images.
This media was adopted by WhatsApp in 2018, nine years after the creation of this app, granting its users the opportunity to create their stickers on the platform. Due to its broad variety of images, stickers allowed users to be more self-represented while also increasing sympathy and social presence~\cite{smileyface_stickers}, therefore becoming a relevant tool to express opinions and feelings through the app.

While there are some efforts dedicated to the study of memes~\cite{mememagic2021cscw,memes_savvas}, few are focused on stickers. These mostly attempt to understand why and how people use stickers on messaging platforms. In \cite{smileyface_stickers} the authors conducted a qualitative approach with semi-structured interviews to understand sticker usage patterns, observing individual and social characteristics of emoticon usage to reinterpret the meaning of emoticons in instant messages. They suggest that, in addition to using stickers to express emotions, users may also have other strategic and functional purposes. Zhou et al.~\cite{emojis_chi} also present a qualitative study of mobile messaging, focusing on the rapid proliferation of emojis and stickers and the decrease in dependence on text within the platform. 

Given the importance of stickers in messaging services and their usage as reaction or responses, some studies go in the direction of predicting a sticker as a reply to a message in a conversation. 
In \cite{stickerrecomendationWWW20}, the authors trained a deep learning model to respond to messages with the most suitable sticker for the situation in conversation, and
\cite{laddha2020understanding} explores the sticker recommendation problem in two steps, the first one being predicting the message that the user is likely to send in the chat and the second one being substituting the predicted message with an appropriate sticker.
Furthermore, the study presented in \cite{whatsapp_stickers} explores the social and cognitive factors that affect the use of stickers on WhatsApp, creating a unified model for the usage and acceptance of stickers in social media messaging.

Stickers can help users define their intentions or feelings in messages, but their complexity and placement constraints can also result in miscommunication. For instance, \cite{stickers_ambiguous} explores the sticker misinterpretations in instant messaging platforms, showing how people perceive emotion in stickers as well as how sticker miscommunication occurs in actual chat contexts. Most of the issues occurred due to stickers' ambiguous (multiple) facial/bodily expressions and different perceptions of dynamism in gestures. In real chats, there were also contextual misinterpretations, that senders and receivers differently interpreted the representation of stickers and the corresponding textual messages.
Another facet of miscommunication is studied by \cite{miscommunication_stickers}. They analyze the outcomes of sticker usage in authentic online communications, finding differences between the sender's intention and the receiver’s interpretation on 34.7\% of the stickers, although they say that these disparities did not adversely affect communication.

It is interesting to observe this miscommunication in sticker-based conversations, especially in the WhatsApp ecosystem, in which misinformation is also an issue on the platform~
\cite{Resende-WWW2019,Bursztyn19thousand}. The app is pointed as a pivot in the spreading of misinformation campaigns during political elections in Brazil and India~\cite{melo2019complexnet}.
Studies on WhatsApp show that the contents of text~\cite{resendewebsci19,machado19misinformation}, image~\cite{Resende-WWW2019,garimella2020images} and audio~\cite{whatsapp_audio,elmasri2022audio} are used to transmit false stories on the platform, leading to the use of systems to monitor these contents\cite{whatsappmonitor@icwsm19}, evidencing a multimodal trait of misinformation within WhatsApp.

Memes are not only used for entertainment, but also carry more in-depth meanings, such as political opinions~\cite{political_meme}. In their work, \cite{memes_savvas} found a substantial number of politics-related memes on both mainstream and fringe Web communities, supporting media reports that memes might be used to promote or harm politicians.
They also observed that Internet memes are increasingly used to sway and manipulate public opinion, and this prompts the need to study their propagation, evolution, and influence across the Web.
Given the parallels between memes and stickers, including that the latter has an even higher emotional link with users, and knowing the important role of WhatsApp in political scenarios, we also need to understand the implications of the usage of this multimedia format within the messaging platforms and its relationship with partisan context.

While many qualitative studies investigate the intention of users in sending stickers, providing meaningful characteristics of this medium and key insights about what they are, we lack a more systematic characterization of this new media format, about how they appear within chat structures, their political implications, and, moreover, the offensive aspect with which they can threaten vulnerable groups. In this work, we fill this research gap by providing a large-scale analysis of the use of stickers in WhatsApp political groups, shedding light on the contribution that stickers have in the architecture of our current digital communication.

\section{Methodology} \label{sec:metho}

Next, we briefly describe out methodology to gather stickers from WhatsApp and process them, as well as some potential limitations of our strategy. 

\subsection{Data Collection}

WhatsApp has more than 2 billion users and more than 100 billion messages are exchanged every day\footnote{\url{https://twitter.com/WhatsApp/status/1364714386078621703}}. Given the closed architecture of the app, this data is mostly covered by end-to-end encryption in private chats. However, there is a large number of public WhatsApp groups shared online from which data can be collected. Through an invite in the form of a URL, any user can join a group and participate in public chats with hundreds of people, whose topics include, but are not limited to, politics, sports, work, and entertainment.
Therefore, an effective and popular practice, then, is to get access to this kind of data by joining these public groups to collect messages shared within them~\cite{garimella2018whatsapp,Resende-WWW2019,whatsappmonitor@icwsm19,Bursztyn19thousand,reis2020dataset}.
In this work, we follow a similar methodology by selecting public groups dedicated to political discussions shared on social networks in Brazil and collecting messages exchanged within these chats during the four months of the Brazilian electoral campaign period, dating back from June of 2022 and up to September of the same year.
For each message, we collect the following data:
(i) the user ID, (ii) the ID of the group in which the message was posted, (iii) the timestamp, (iv) whether the message was labeled ``forwarded'' or not, (v) the text of the message, (vi) the media type of the message (i.e. sticker, image, audio, video or document) and, when available, (vii) the actual attached multimedia file. In Table~\ref{tab:dataset} we show the summary of the data collected. In total, we collected more than 6M WhatsApp messages, of which 659K were stickers sent in 1,063 different groups by more than 64k users.

\begin{table}[t]
    \centering
    \caption{Summary of data collected from WhatsApp.}
    \label{tab:dataset}
    \resizebox{\columnwidth}{!}{%
        \begin{tabular}{ccccc}
            
            \toprule
            \textbf{\begin{tabular}[c]{@{}c@{}}Number of \\ Messages\end{tabular}} & \textbf{\begin{tabular}[c]{@{}c@{}}Number of  \\ Stickers\end{tabular}} & \textbf{\begin{tabular}[c]{@{}c@{}}Unique \\ Stickers\end{tabular}} & \textbf{\begin{tabular}[c]{@{}c@{}}Number of \\  Groups\end{tabular}} & \textbf{\begin{tabular}[c]{@{}c@{}}Number  \\ of Users\end{tabular}} 
            \\ \midrule
            6,039,760                                                          & 659,576  (10.9\%)                                                  & 57,031                                                              & 1,063                                                            & 64,227                                                          
            \\ \bottomrule
        \end{tabular}
    }
\end{table}

\subsection{Processing Sticker Data}

After collecting the data, we filter messages by media type, selecting all stickers sent in the chats. For each individual sticker, we also processed a visual hash for the image file related to it\footnote{Note that, for animated stickers, we convert them to a static one using the first frame so that it can be processed as an image}. For this step, we use the pHash algorithm, which takes as input the sticker file and whose output is a 64bit-integer hash, calculated using a discrete cosine transform that extracts a specific ``fingerprint'' to each file~\cite{zauner2010phash}. As perceptual hashes are constructed in a way that small variations in the pixels of the input image will result in similar hashes (opposing to crypt hashes, in which a single byte difference will drastically change the final hash), this process allows us to visually compare two similar images~\cite{memes_savvas}. By processing all 659K sticker messages, we found 57,031 unique stickers (i.e., with exactly the same hash).
For each sticker, we also extract an attribute of its dominant color by grabbing the main color from the palette of that image\footnote{\url{https://pypi.org/project/colorthief/}}. After that, we represent each sticker file as a combination of its hash and color. 

We use this representation of the images to cluster them together and obtain insights that emerge from the clusters. A similar methodology was used to study memes from Web communities~\cite{memes_savvas}, in which annotated images were mapped to their corresponding clusters. Using the DBSCAN algorithm, we can provide an investigation on the popularity and diversity of stickers on WhatsApp and how users make use of this form of media.

Besides the visual appearance of stickers, we also assess them by the context in which they are used. For that, we grouped stickers based on the group network connection in which they were sent by building a graph of the stickers sent in the same group. We also labeled the political groups whose data were collected as right- or left-wing alignments to investigate the relationship between the graph topology and the political alignment of their nodes.

Finally, we analyzed the abusive use of stickers on WhatsApp. 
We use the NSFW Yahoo model\cite{nsfw_yahoo} to measure information of how ``explicit'' each sticker is. We also use the SafeSearch detection with Google Vision API\footnote{\url{https://cloud.google.com/vision/docs/detecting-safe-search}} to get complementary data about potentially harmful content being posted on WhatsApp through the stickers.

\noindent 
\textbf{Data Limitations.} 
It is known that, at least for WhatsApp, most of the conversations occur in private chats. Therefore, this data collection may be not representative of the entire platform group dynamic, especially considering that we observe only public groups whose topics are dedicated to political discussion. Nevertheless, this methodology collects a large volume of data from the app Since political groups are particularly inclined to massively share content on WhatsApp, they can bring meaningful insights for the discussion. It is important to highlight, however, that results refer only to content circulating on the public layer of the platform and within the context of politics, so they may not hold true for the entire network. Despite this limitation, our efforts aim to shed light on important aspects that can support others towards understanding how the WhatsApp platform is being abused, since some of these aspects would be otherwise hidden behind the closed nature of the platform's architecture.

\vspace{0.05in}
\noindent 
\textbf{Ethical considerations.} Unlike other approaches, this work does not involve direct experiments with human subjects. Regarding the data analyzed, we collected messages only from publicly accessible groups on WhatsApp and in a public political context. In addition, we look exclusively at stickers sent within those chats, and all sensitive information (i.e. usernames, phone numbers, and further text messages) has been previously anonymized to ensure the privacy of users. Last, we are providing a large-scale study regarding the use of stickers in WhatsApp political groups presenting the results in aggregated way.


\section{Stickers as a distinctive form of media} \label{sec:results1}

Multimedia messages are very popular on WhatsApp~\cite{Resende-WWW2019}. Even though text messages still make up the majority of the data, users often share different kinds of media using images, videos, audio, and stickers. Although images and stickers are very similar (both are visual media displayed in the middle of the chat) some key changes distinguish them. Stickers may show an animated image on chat, they are smaller than actual images, they are stored in a different format by the app and they are usually displayed directly to users on chat (as emojis) while images need do be downloaded.
These characteristics, however, refer to the structure of the sticker media and not about their usage. We do not know whether the thought process employed by users when sending a sticker is similar or not to other kinds of media. In this section, we evaluate some aspects of the stickers to investigate how members of political groups use this medium.

\begin{figure}[t]
 \centering
 \subfigure[Groups]{\includegraphics[trim={0 0 0 0},clip,width=0.75\linewidth]{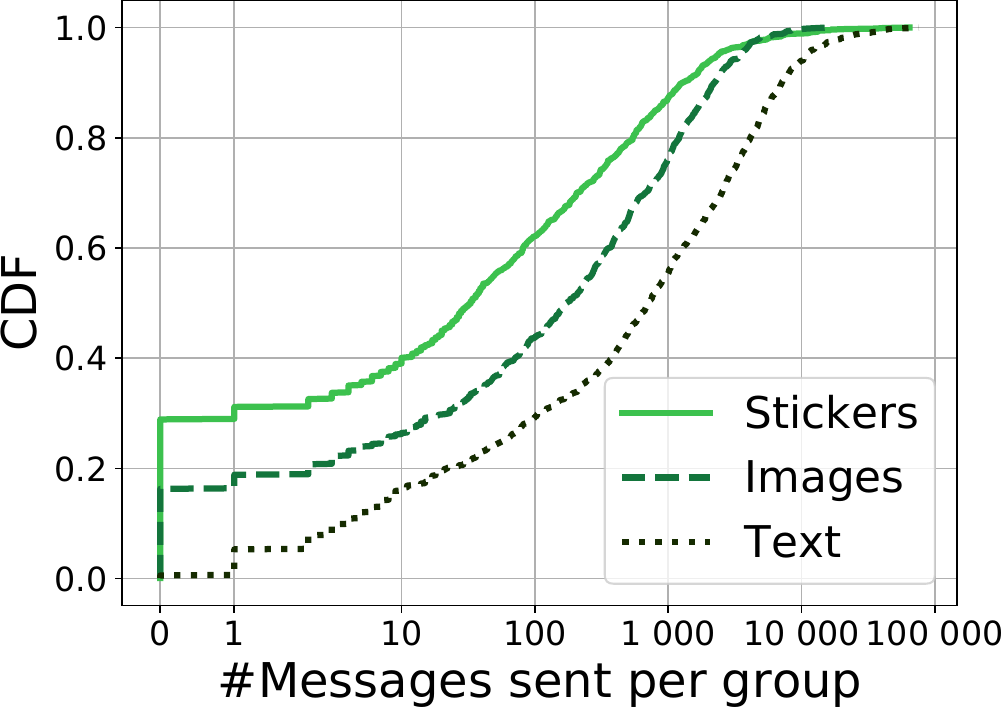}\label{fig:cdf_groups}}
 \hfill
  \subfigure[Users]{\includegraphics[trim={0 0 0 0},clip,width=0.75\linewidth]{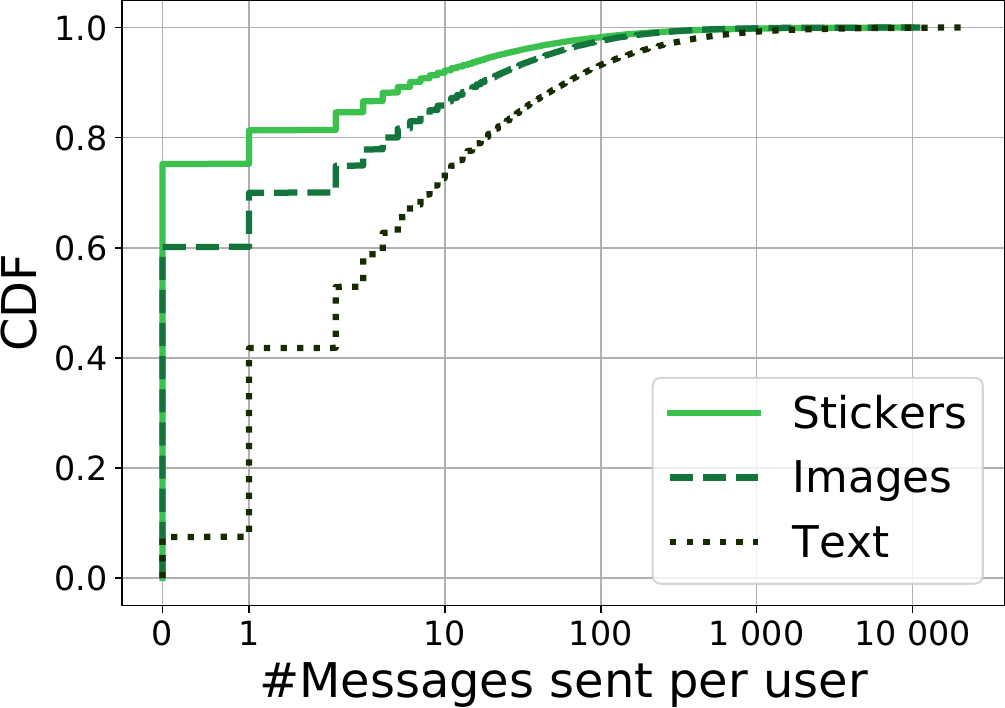}\label{fig:cdf_users}}
  \caption{Cumulative Distribution Function (CDF) of stickers sent per groups and users compared to image and text.}
  \label{fig:cdf_totalmessages}
\end{figure}

In Figure~\ref{fig:cdf_totalmessages}, we compare stickers with images and texts regarding the volume of media sent per group and per user. In the cumulative distribution function (CDF) of the messages per group (Fig.~\ref{fig:cdf_groups}), we see that sticker messages were less popular in the groups than images. About 30\% of the groups did not send any stickers in the chat; images were absent in 20\% of them. Moreover, 60\% of groups sent a maximum of 100 stickers, while 40\% of the groups sent less than that. Still, we observe that some groups have more than 10k stickers and images.
In the distribution of messages per user (Fig.~\ref{fig:cdf_users}), we have similar curves, with stickers being the least common medium. Here, even more users do not send any stickers (75\%), but we also have users who individually sent around 10k stickers.
Hence, we can observe that, as expected in a messaging app, text format is much more used than other types of media. However, it is interesting to see that users prefer to send images than stickers on the groups, even though both media are widely spread on WhatsApp, especially when considering that stickers are generally easier to send, as they do not need to select a file from the gallery and are more incorporated into the WhatsApp interface.


\begin{figure}[t]
    \centering
    \subfigure[Shares]{\includegraphics[width=0.75\linewidth]{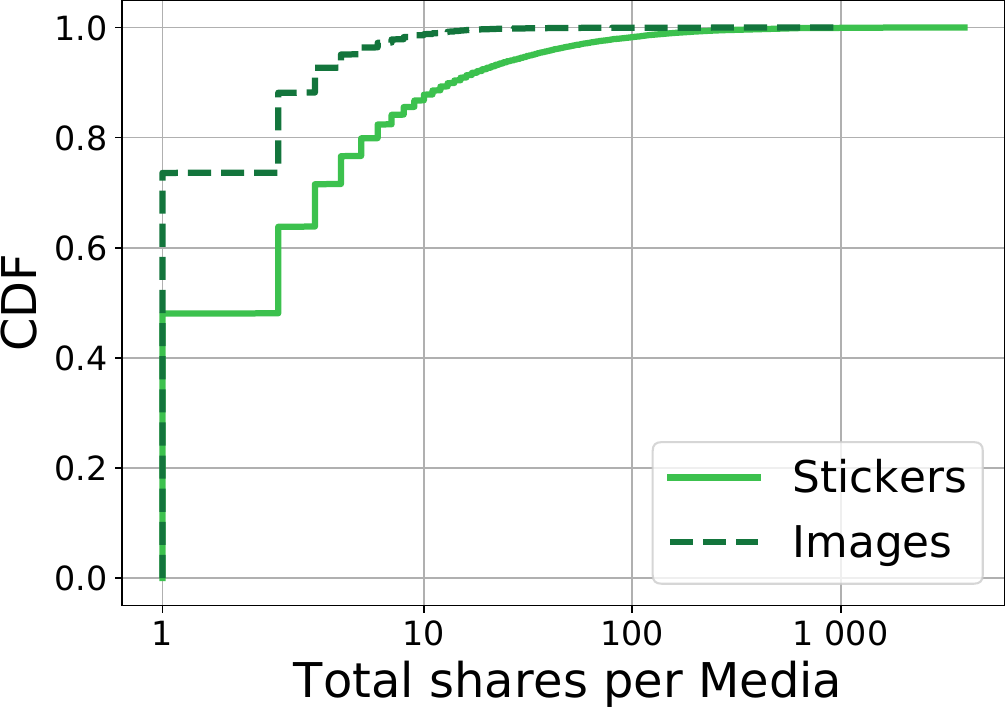}\label{fig:cdf_shares}}
    \hfill
    \subfigure[Forwards]{\includegraphics[width=0.75\linewidth]{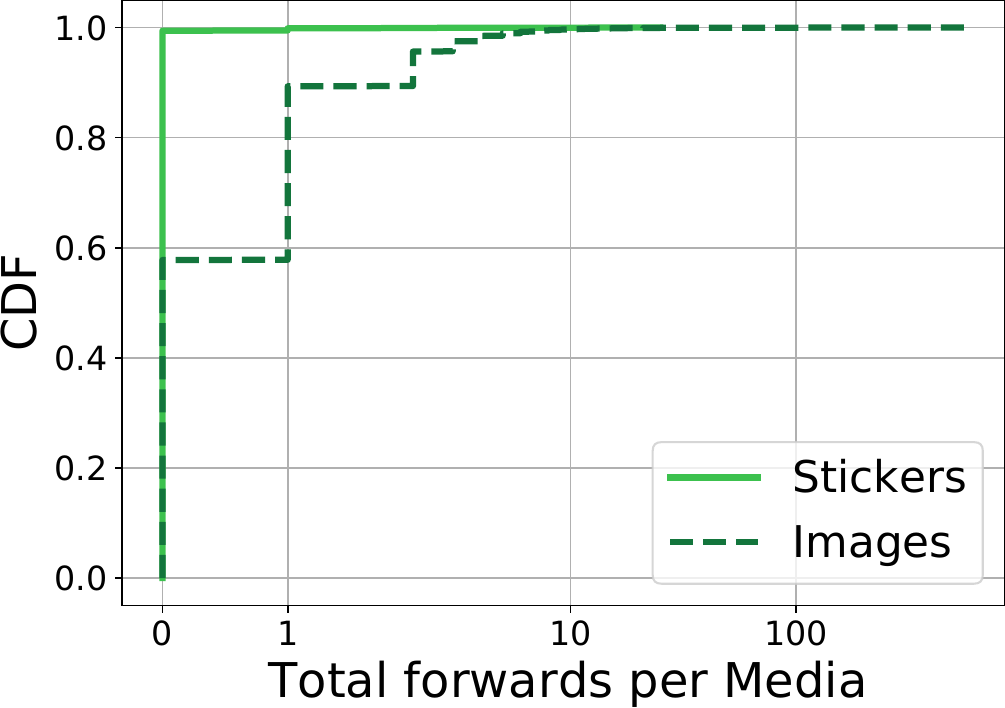}\label{fig:cdf_forwards}}
    \caption{Cumulative Distribution Function (CDF) of total shares and forwarding per sticker and image medias.}
    \label{fig:cdf_mergedmedia}
\end{figure}

As we processed and merged stickers, we measured the number of shares each sticker has within our dataset. In Figure~\ref{fig:cdf_shares} we evaluate the distribution of total shares per media by comparing images and stickers. Even though we observed more image messages in the data, when looking at the total shares per unique media, we note that stickers are more shared than images. About half of the stickers appeared more than once, many of which were shared more than a thousand times. In contrast, about 75\% of images have only a single appearance.

Another attribute we obtained for each message is whether it was forwarded or not. Forwarding is an important tool in WhatsApp communication, in which users share content with their contact~\cite{whatsapp_forwarding}. When available, this info allows us to determine how many messages were forwarded for each piece of media in our dataset.  
By analyzing this quantity, we found that images are much more frequently forwarded than stickers. Almost none of the stickers collected were found in forwarded messages, meaning that nearly all of them were sent directly from the associated users to a group. On the other hand, for images, we observe that 40\% have been shared as a forward at least once. 
We can see that stickers not only have a different structure when compared to images, but they also present a different usage by the users. In general, there are fewer stickers than other media in groups, and users also send fewer stickers. However, they are individually shared more often and directly by user within chat. A possible explanation for this is that stickers can have a collectible tone \cite{emojis_chi}, in which users prefer to save and keep a collection to have at their disposal a range of options to use in a specific situation or as a self-disclosure reaction in the chat~\cite{wang2016more} instead of just forward a piece of content as they do for images. This implies that a sticker can be reproduced more than once in different scenarios, while images are usually more concise and portray a specific piece of information.

\section{Stickers Characterization} \label{sec:results2}

In this section, we present different forms of grouping stickers, including by visual resemblance using the pHash and by context through building the graph of stickers according to the groups in which they were posted.

\begin{figure}[t]
    \centering
    \includegraphics[trim={0 0 0 0},clip,width=1.0\linewidth]{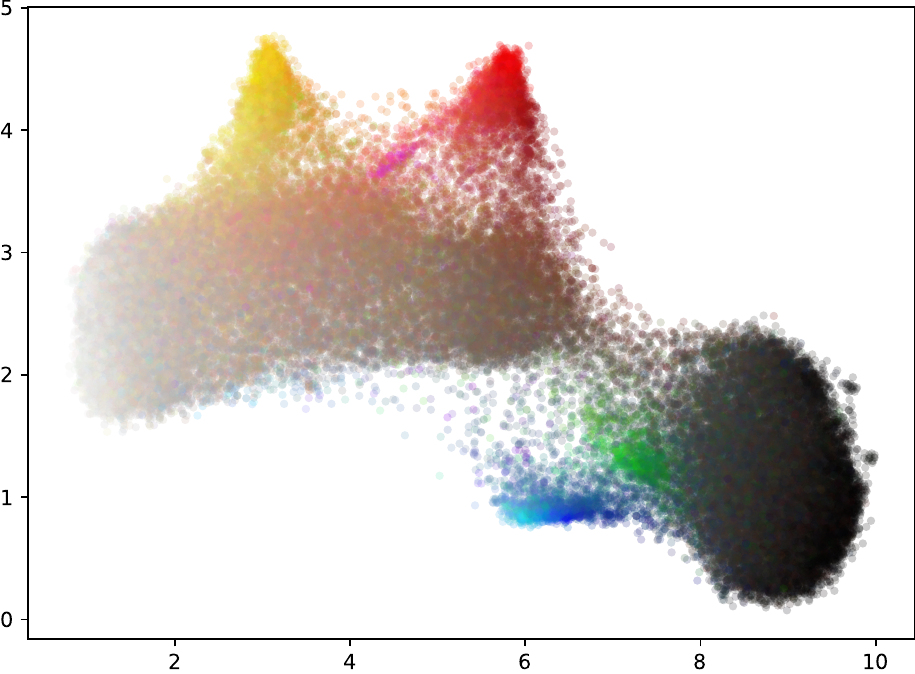}
    \vspace{-0.2cm}
    \caption{UMAP visualization of all stickers from WhatsApp dataset.}
    \label{fig:UMAP}
\end{figure}


A visual representation of all stickers collected is shown in Figure~\ref{fig:UMAP}. We create a depiction by using both the pHash binary vector and each sticker's dominant color. Then, we employ dimensionality reduction via UMAP to plot the sticker representations in a 2D space. Each point is associated with a single sticker's representation and its color is the actual sticker's dominant color. It is visible that there is a concentration of red and also green and blue stickers, which are related to the political context we monitored, and many stickers within the data are directly related to political campaigns for the 2022 Brazilian elections. The leading left-wing party in Brazil employs red colors associated with the party's branding, while the leading nationalist right-wing party mostly uses green, blue, and yellow, colors derived from the Brazilian flag. The stickers in our dataset reflect the partisan nature of the branding of the competing parties. A thorough manual investigation of the yellow chunk of stickers revealed presence of many ``emoji'' stickers, which are faces drawn in emoji style representing existing emoji characters. The grayish area in the plot encompasses many different sticker styles and we do not could point to a single characteristic of them.

\subsection{Clustering by Visual Similarity}

To explore the similarity of these stickers, we leverage features of images to cluster them together and investigate whether there exist patterns and characteristics to be observed within stickers on WhatsApp. A similar methodology was employed for the study of memes in Web communities~\cite{memes_savvas}, annotating and mapping them to clusters.  
Since the perceptual hash (pHash) offers a comparable fingerprint of stickers, we can use it to cluster our stickers through the density-based spatial clustering of applications with noise (DBSCAN) algorithm, merging them in visually similar groups of content shared on WhatsApp. Using DBSCAN enables the investigation of the popularity and diversity of stickers found on WhatsApp and insights into how users may use this form of media.

\begin{figure}[t]
  \centering
  \subfigure[]{\includegraphics[trim={0 0 0 0},clip,width=0.24\linewidth]{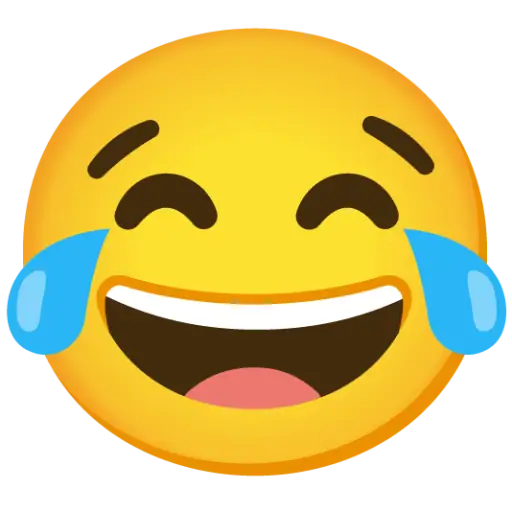}}
  \subfigure[]{\includegraphics[trim={0 0 0 0},clip,width=0.24\linewidth]{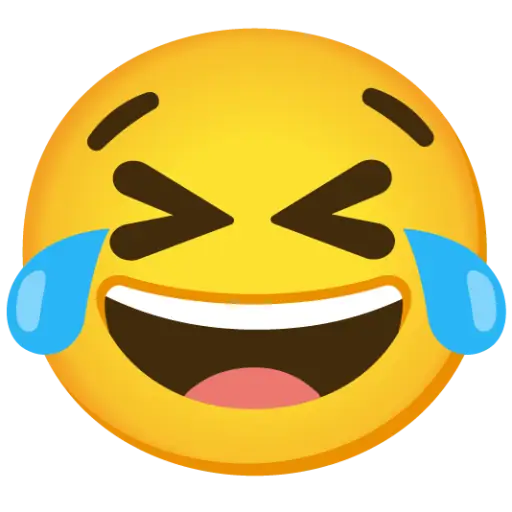}}
  \subfigure[]{\includegraphics[trim={0 0 0 0},clip,width=0.24\linewidth]{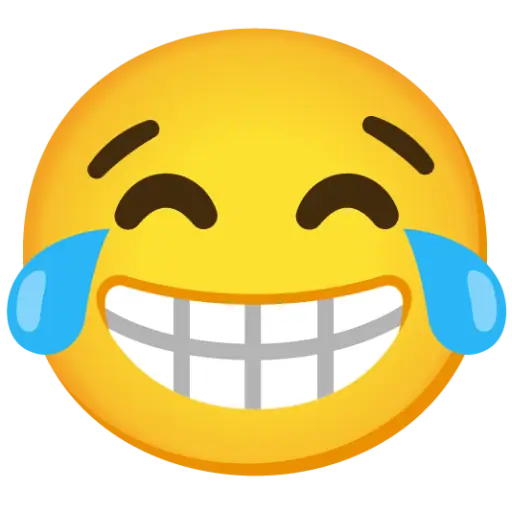}}
  \subfigure[]{\includegraphics[trim={0 0 0 0},clip,width=0.24\linewidth]{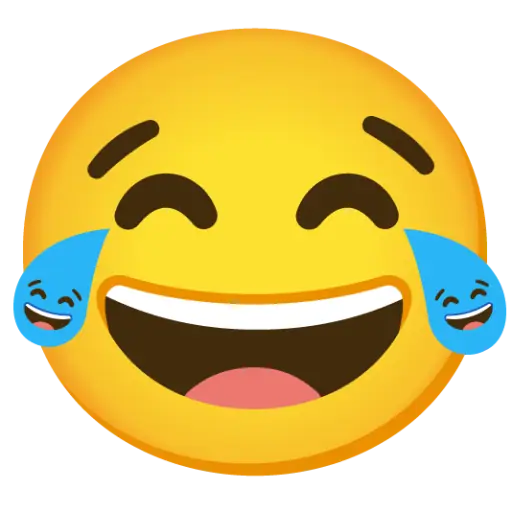}}
  \caption{Example of a cluster of stickers representing emojis.}
  \label{fig:emojis}
\end{figure}

\begin{figure}[t]
    \centering
    \subfigure[]{\includegraphics[trim={0 0 0 0},clip,width=0.23\linewidth]{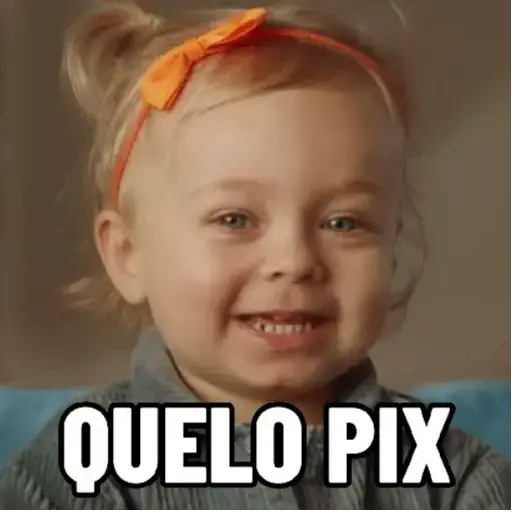}}
    \subfigure[]{\includegraphics[trim={0 0 0 0},clip,width=0.23\linewidth]{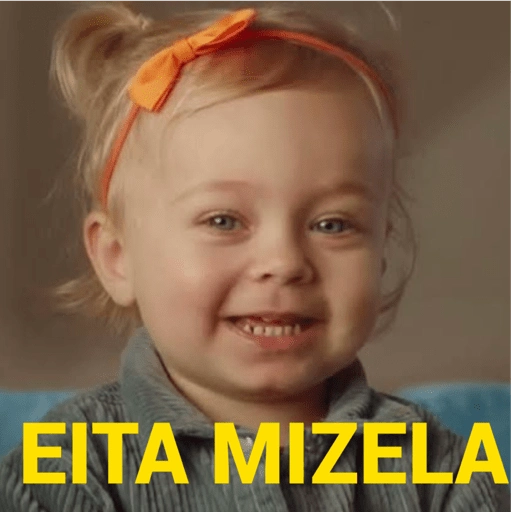}}
    \subfigure[]{\includegraphics[trim={0 0 0 0},clip,width=0.23\linewidth]{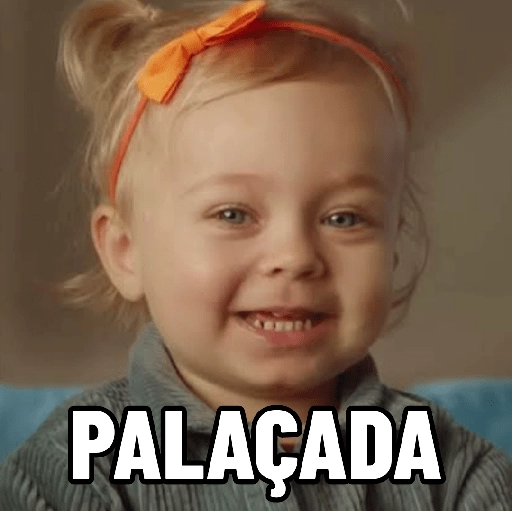}}
    \subfigure[]{\includegraphics[trim={0 0 0 0},clip,width=0.24\linewidth]{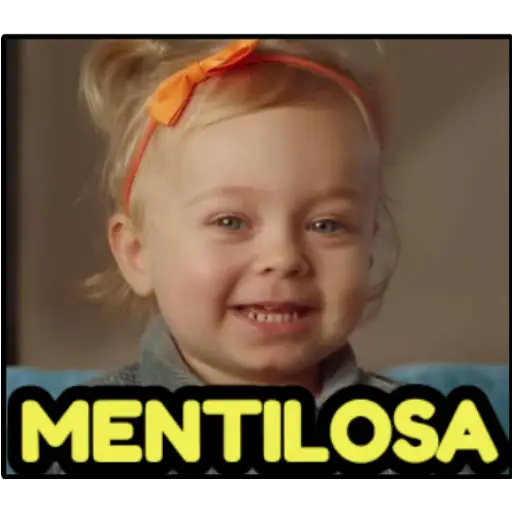}}
    \caption{Example of the cluster with meme sticker template with small variations.}
    \label{fig:meme}
\end{figure}

Among the ensuing clusters, we find pHash-grouped stickers that represent variations of a base image, emoji, or meme. 
Some of the resulting clusters demonstrate a key aspect of stickers: their similar use as a meme and also as emojis, as shown in Figures~\ref{fig:emojis} and ~\ref{fig:meme}. 
Emojis are text characters that refer directly to pictorial representations of feelings (often in the form of round yellow faces) and
other objects~\cite{emojis_chi}. There are clusters with elaborated versions of those characters, occasionally even as animated stickers. These are usually employed as reactions to other messages, in a similar fashion emojis appear in text messages ~\cite{wang2019culturally}.
The clusters associated with meme templates put together analogous images that share a unique base template, but are differentiated by slight changes in text or detail to instantiate a joke or a funny situation, but are still recognized as the same meme~\cite{ge2020memeticstickers}. This pattern is also observed in memes, in which popular standardized comical images are widely spread across the Internet.  This behavior of creating, remixing, transforming, and editing variations of an image is a common practice in meme distribution on the Web~\cite{smileyface_stickers,mememagic2021cscw,davison2012language} 

\begin{figure}[ ]
  \centering
  \subfigure[]{\includegraphics[trim={0 0 0 0},clip,width=0.35\linewidth]{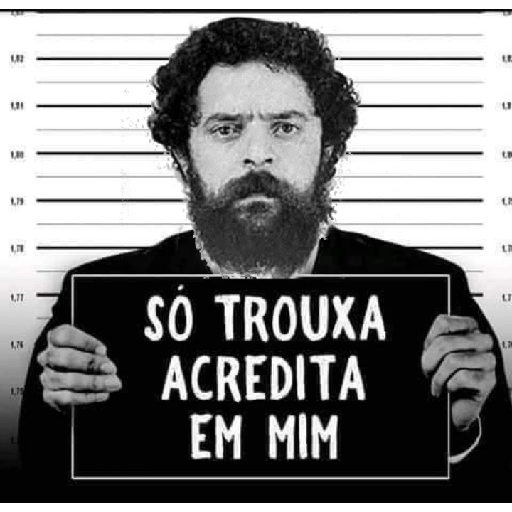}}
  \subfigure[]{\includegraphics[trim={0 0 0 0},clip,width=0.35\linewidth]{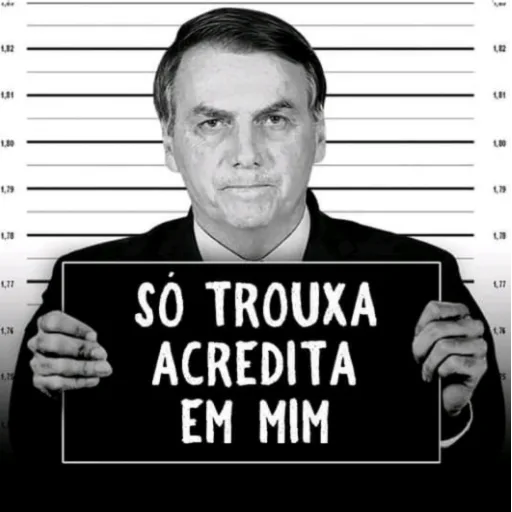}}
  \caption{Examples of visually similar stickers used by opposing political leanings.}
  \label{fig:opposition}
\end{figure}

It is interesting to note that within certain clusters, very visually similar stickers may also portray opposing ideas.
In the context of the public political groups encompassed in this work, many clusters are related to the poll-leading politicians in the Brazilian 2022 elections and contain advertising/provocative material associated with their campaigns. Here, we can find examples of corresponding sticker images that carry drastically opposing partisan content. Figure~\ref{fig:opposition} offers an example of a cluster that grouped two visually analogous stickers but which are shared in different contexts. The stickers portray an edited image of opposing candidates of the left and right-wing parties in a criminal mugshot, suggesting that stickers are coopted by adversary groups and used with opposing semantic but adopting with the same visual aesthetics.

These results alert us to the implications of grouping stickers exclusively based on their visual appearance. Small details oftentimes imply drastic changes in the semantic value of the sticker. 


\subsection{Political Alignment of Stickers}

\begin{figure}[t]
    \centering
    \includegraphics[trim={0 0 0 0},clip,width=1.0\linewidth]{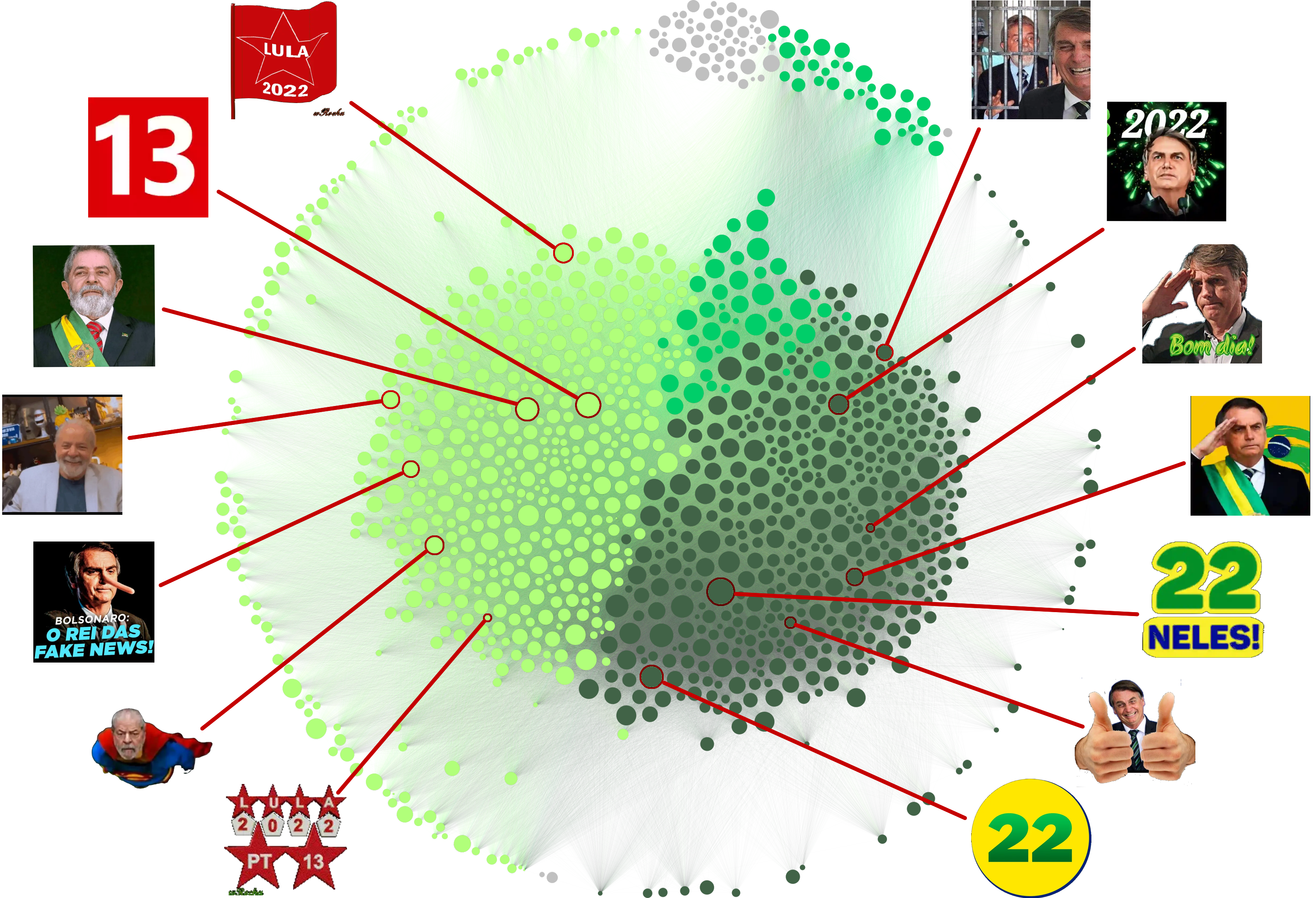}
    \caption{Sticker network.}
    \label{fig:sticker_network}
\end{figure}

As exemplified in the previous section, even though stickers can be very visually similar, they might contain drastically different ideas and thus be used in very distinct scenarios. To evaluate the usage of stickers from this perspective, we built a network of stickers based on the groups in which they were shared.  Figure~\ref{fig:sticker_network} offers a graph visualization of this network assembled with the most popular stickers in our WhatsApp data. 
There, each node is a sticker, and they are linked with an edge if both appear in the same group, we also applied a community detection algorithm based on modularity optimization~\cite{community_detection} to extract the communities of stickers in this network. 
For visualization clarity, we used only stickers shared more than 100 times in the observed groups and edges with at least 10 groups in common.


Even with a very dense network (density measured at 0.698), we could identify two main larger communities and a third smaller one.  These two main communities reflect the polarized nature of our data and have very analogous stickers from both political leaning.  
There were two major candidates for the Brazilian 2022 presidential elections (Bolsonaro, current right-wing Brazilian president, and Lula, former leftist president of Brazil). The majority of the stickers found on the monitored public groups, therefore, represent these sides of this dispute. Some stickers are symbols, numbers, and colors associated with each political party, while others depict their candidate if choice as a superhero or sporting items associated with the elected president of Brazil. Numerous stickers aim to criticize or enrage members of the opposing political leaning. Together, these sticker archetypes form a rather unique and connected political community. The smaller third community, however, presents a more diverse type of content, including some with sexually explicit nature.

To evaluate the partisan leaning of stickers, we label each monitored group as left-wing, right-wing, or undefined, following the political stance declared by the group's name and, if necessary, by their description. The annotation shows that most groups support the right-wing candidate Bolsonaro, with 666 groups, whereas just 126 groups support a left-wing agenda. The remaining 617 groups were tagged as undefined as it was not possible to determine their alignment (e.g. groups of support of a different political leaning or for general political debate).

Next, we examine each sticker to determine its political leaning based on the groups that posted them. Then, taking into account the label of these groups, we calculate the proportion associated with each political alignment. Finally, we assigned a political alignment label for the stickers according to the side of the political spectrum that had the highest portion of messages. This label indicates if the sticker is mostly used in the context of right- or left-wing groups. Through this process, we characterize the political nature of stickers shared through Brazilian WhatsApp groups.
Note that certain stickers might be equally used by both sides of the political spectrum, or also predominantly by groups that are not clearly defined politically. To prevent mislabeling in such cases, we introduced a third category called ``undefined'', used for stickers that do not have a 60\% majority on either side. This approach allows us to better identify stickers that are closely linked to political discussions. Through this annotation, we found that 18k stickers are biased towards usage in right-wing groups and 6k towards the left, as shown in Table~\ref{tab:political}.  Note that the difference could be explained by the fact that, in Brazil, the right-wing movement is seen to be actively mobilized on the WhatsApp platform~\cite{Resende-WWW2019,Bursztyn19thousand}.

\begin{table}[]
    \centering
    \caption{Political alignment of groups and sticker context.}
    \label{tab:political}
    \begin{tabular}{lccc}
        \toprule
                                             & \textbf{Right} & \textbf{Left} & \textbf{Undefined} \\ 
        \midrule
        \multicolumn{1}{l}{Total Groups}   & 666   & 126  & 617       \\ 
        \multicolumn{1}{l}{Total Stickers} & 18,158 & 6,704 & 32,169     \\ 
        \bottomrule
    \end{tabular}
\end{table}

Given the potential for stickers to convey strong partisan sentiments, our investigation delves into the hypothesis that political stickers serve as tools for provocation or even direct attacks against opposing political groups. The public nature of the monitored groups allows individuals from adversarial political positions to join, establishing a channel for launching targeted assaults against individuals aligned with opposing ideologies. Leveraging the ease of sticker sharing with just a single click, these initiatives are particularly conducive to flooding group chats with divisive content. To examine this phenomenon, we explore whether stickers predominantly associated with a specific political alignment also make appearances within groups representing the opposing political spectrum.

Our analysis reveals that the majority of politically annotated stickers were, in fact, exclusively utilized within their corresponding ideological circles. However, we also identified instances where political stickers transcended partisan boundaries, being shared in groups with opposing political affiliations. Employing our criteria, we identified 869 partisan stickers, with a notable majority (794) exhibiting right-wing bias and being disseminated within leftist groups. In contrast, only 75 left-wing biased stickers were observed in right-wing groups.

\begin{figure}[t]
  \centering
  \subfigure[]{\includegraphics[trim={0 0 0 0},clip,width=0.23\linewidth]{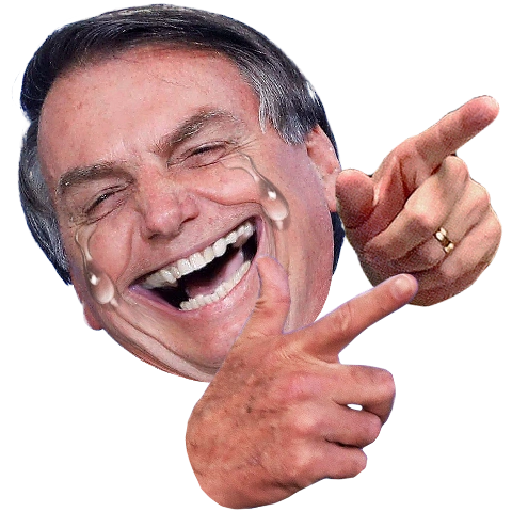}\label{fig:political10}}
  \subfigure[]{\includegraphics[trim={0 0 0 0},clip,width=0.23\linewidth]{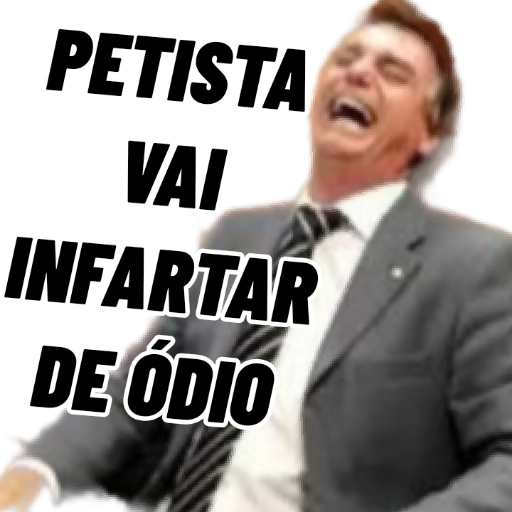}\label{fig:political1}}
  \subfigure[]{\includegraphics[trim={0 0 0 0},clip,width=0.23\linewidth]{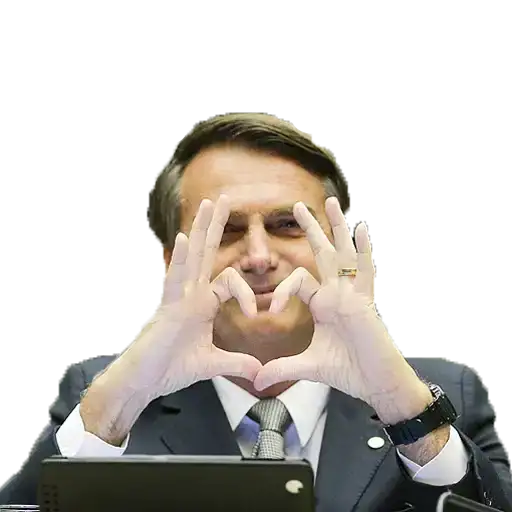}\label{fig:political2}}
  \subfigure[]{\includegraphics[trim={0 0 0 0},clip,width=0.23\linewidth]{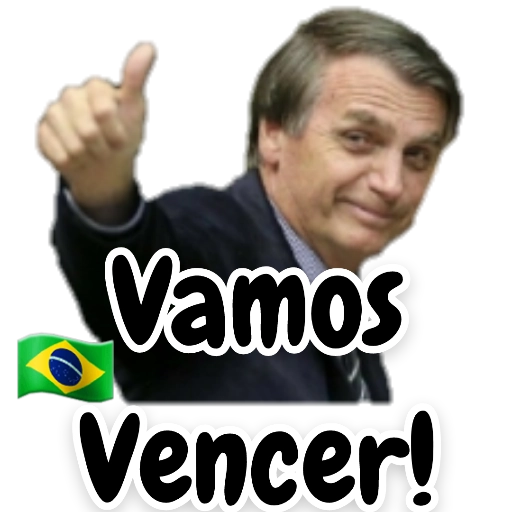}\label{fig:political3}}
  \subfigure[]{\includegraphics[trim={0 0 0 0},clip,width=0.23\linewidth]{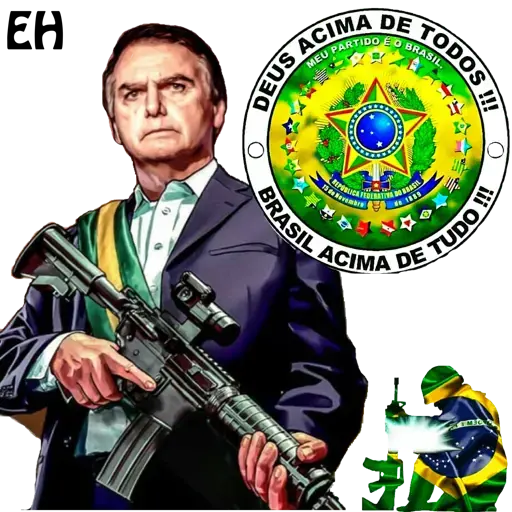}\label{fig:political6}}
  \subfigure[]{\includegraphics[trim={0 0 0 0},clip,width=0.23\linewidth]{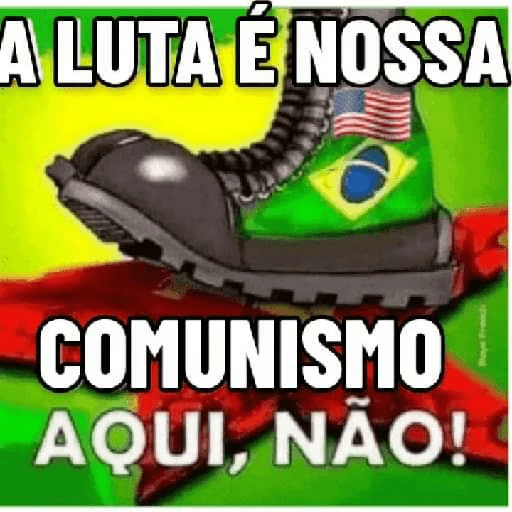}\label{fig:political8}}
  \subfigure[]{\includegraphics[trim={0 0 0 0},clip,width=0.23\linewidth]{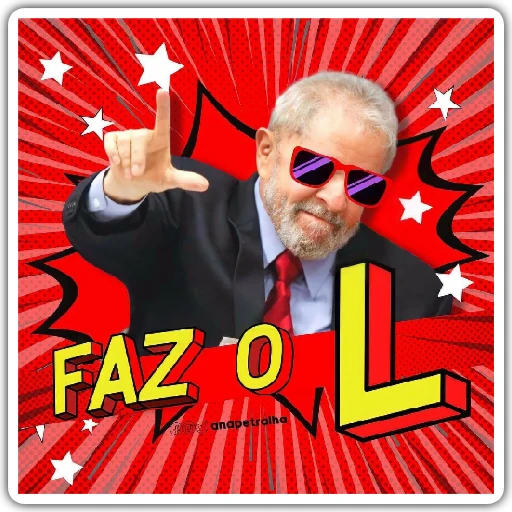}\label{fig:political4}}
  \subfigure[]{\includegraphics[trim={0 0 0 0},clip,width=0.23\linewidth]{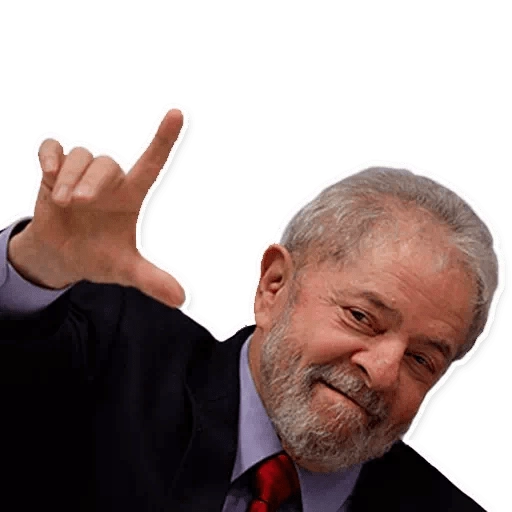}\label{fig:political5}}
  \caption{Stickers used to ``provoke'' opposing political groups on WhatsApp.}
  \label{fig:political}
\end{figure}

Figure~\ref{fig:political} showcases examples of these ``politically inconsistent'' stickers, which promote a partisan candidate in rival environments. The presence of such stickers in politically opposing groups suggests a deliberate act of confrontation by users, trying to cause reactions and distress among individuals holding contrasting political views. For instance, Figures \ref{fig:political10},\ref{fig:political1},\ref{fig:political2},\ref{fig:political3} depict stickers featuring the right-wing presidential candidate Bolsonaro, perceived as provocative within leftist groups. One sticker even includes the text ``Leftists will infarct with hate'', tailored to instigate distress. Similarly, leftist stickers depicted in Figures~\ref{fig:political4} and \ref{fig:political5} endorse Lula and are incongruous with right-wing ideologies.
Of particular concern are images such as those in Figure~\ref{fig:political6}, depicting Bolsonaro wielding a firearm, and Figure~\ref{fig:political8}, featuring heavy boots crushing a red star with the words ``This is our fight. Communism, not here!'' These images evoke menacing connotations, symbolizing a dominance of right-wing ideologies over leftist counterparts, thereby exacerbating tensions within a polarized online political discourse.

Our findings reveal that clustering stickers based on their image pHashes, which denotes visual similarity, enables the identification of image sets exhibiting even minor variations. However, this may not capture the subtle political context carried with them and may also inadvertently put together stickers with opposite political messages. On the other hand, examining stickers through the perspective of the groups in which they are shared, and thus their political alignment, exhibits a strong association with the contextual environment rather than solely relying on visual resemblances of the images. This highlights the importance of considering the broader socio-political context in understanding sticker usage patterns.

Additionally, we observed that stickers often demonstrate remarkably partisan alignments and are frequently utilized as tools for provocation and abuse within political public groups on WhatsApp. This highlights the pivotal role of stickers as a media for expressing and perpetuating political ideologies and tensions within this network.

Moreover, the asymmetric distribution of stickers aligned with right-wing ideologies may contribute to the reinforcement of echo chambers and polarization within WhatsApp groups. Users who predominantly encounter content aligned with their own political beliefs may become further entrenched in their viewpoints, hindering constructive dialogue and fostering division within online communities. Also, the mechanism of sharing biased stickers within WhatsApp raises questions about the dynamics of political engagement and manipulation on social media platforms. The deliberate dissemination of partisan content, particularly in the form of stickers designed to provoke or attack opposing political groups, shows a strategic usage of digital media for political propaganda and agitation.

In conclusion,  the prevalence of right-wing biased stickers in our dataset shed light on the complex interplay between digital media, political discourse, and ideological polarization on WhatsApp in Brazil. Understanding this ecosystem is essential for promoting informed and inclusive dialogue within online communities and mitigating the risks of manipulation and division.

\vspace{-0.2cm}

\section{Abusive Stickers on WhatsApp} \label{sec:results3}

While political attacks are one domain through which stickers can be abusive~\cite{whatsappattacks_icwsm24}, there are also stickers disseminating offensive and inappropriate content on WhatsApp. In this section, we delve into these pathways, shedding light on the various forms of abusive behavior facilitated by stickers on the platform.


On WhatsApp, users can freely create their own stickers based on any image they want. However, this model is not unanimous among all messaging apps. Most of them have only predetermined and limited sets of stickers that users can use during conversations. Facebook and its messaging system ``Messenger'' does not allow all users to add their own creations as stickers~\cite{facebook_creation}. The mobile messaging app LINE even sells millions of curated sticker packs per month, and stickers have long been one of LINE’s key revenue drivers~\cite{line_revenue}. 
Despite Meta restricting stickers on its other platforms and selling stickers can be a highly profitable economic model, for WhatsApp particularly, Meta chose to let users freely create and share their own stickers. However, this permissive model comes with some challenges, as it opens the door for the dissemination of offensive and inappropriate content.

\begin{figure}[ ]
  \centering
  \subfigure[]{\includegraphics[trim={0 0 0 0},clip,width=0.24\linewidth]{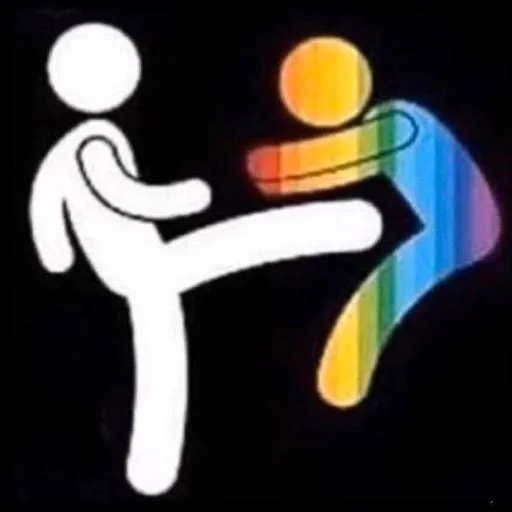}\label{fig:homophobic}}
  \subfigure[]{\includegraphics[trim={0 0 0 0},clip,width=0.24\linewidth]{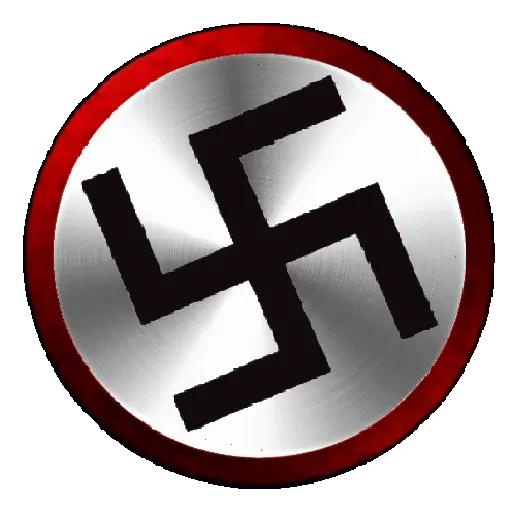}\label{fig:nazist}}
  \subfigure[]{\includegraphics[trim={0 0 0 0},clip,width=0.24\linewidth]{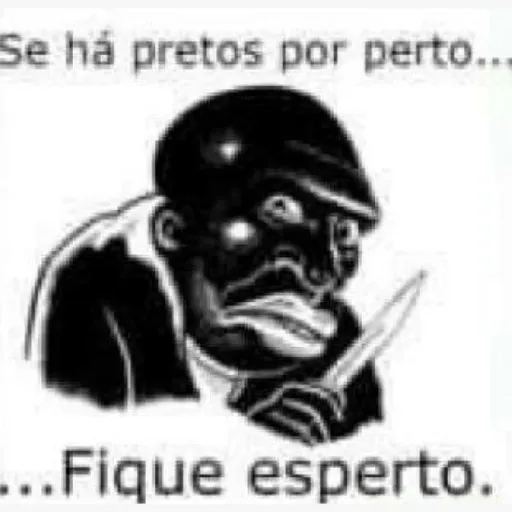}\label{fig:racist}}
  \subfigure[]{\includegraphics[trim={0 0 0 0},clip,width=0.24\linewidth]{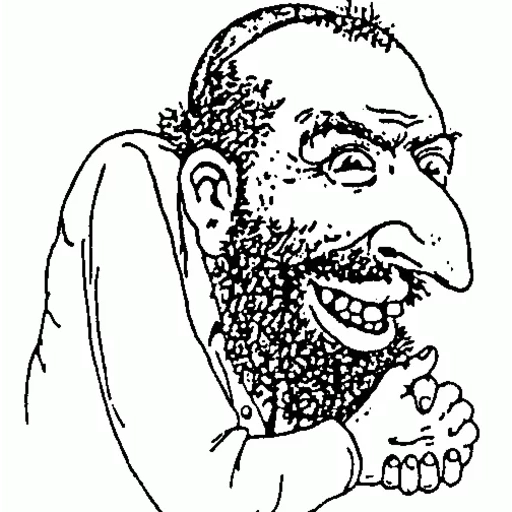}\label{fig:jews}}
  \caption{Example of offensive stickers sent by users that violate terms of use of WhatsApp.}
  \label{fig:offensive_stickers}
\end{figure}

WhatsApp's terms of service, however, determine that stickers created must be legal, authorized, and acceptable images. Furthermore, users are not allowed to use WhatsApp services in ways ``\textit{that are obscene, defamatory, threatening, intimidating, harassing, hateful, racially or ethnically offensive, or instigate or encourage conduct that would be illegal, such as promoting violent crimes}''\footnote{\url{https://www.whatsapp.com/legal/terms-of-service}}.
Although WhatsApp's TOS does not allow for offensive stickers, we find clear examples of stickers that do not comply with these rules. 
Figure \ref{fig:offensive_stickers} shows unacceptable stickers found within our WhatsApp dataset that are highly offensive towards specific groups and violate the terms of use of WhatsApp. 
Sticker˜\ref{fig:homophobic} is an example of a homophobic image against LGBTQIA+ people; Sticker˜\ref{fig:nazist} depicts a swastika, a symbol amply associated with the former Nazi party;  Sticker~\ref{fig:racist} depicts a black male holding a knife and the phrase ``around blacks never relax'', which is a recurring racist remark on the Web~\cite{black_meme}. Similarly, Sticker~\ref{fig:jews} portrays a derogatory illustration of a Jewish man based on antisemitic characterizations of Jewish people, which is also a widely known hate symbol~\cite{jews_meme}.
These examples demonstrate that stickers, uploaded to and freely shared between users on WhatsApp, can be used as media for harassment and targeted attacks toward vulnerable and marginalized people.

\begin{figure}[t]
    \centering
    \subfigure[NSFW Score]{\includegraphics[width=0.75\linewidth]{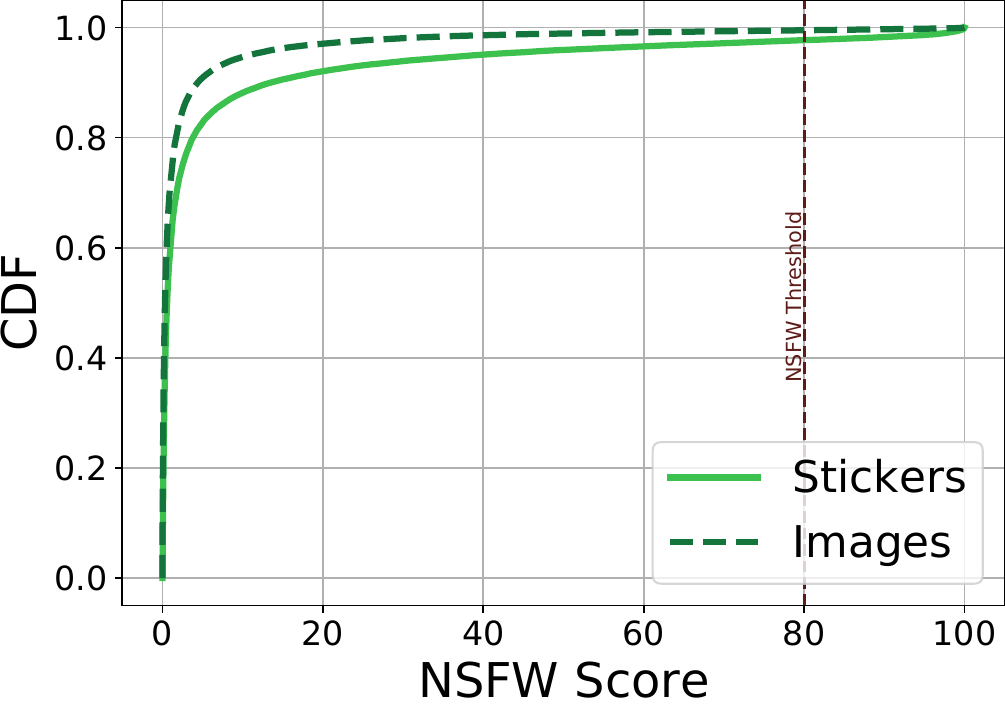}\label{fig:nsfw_stickers_images}}
    \hfill
    \subfigure[NSFW per Users]{\includegraphics[width=0.75\linewidth]{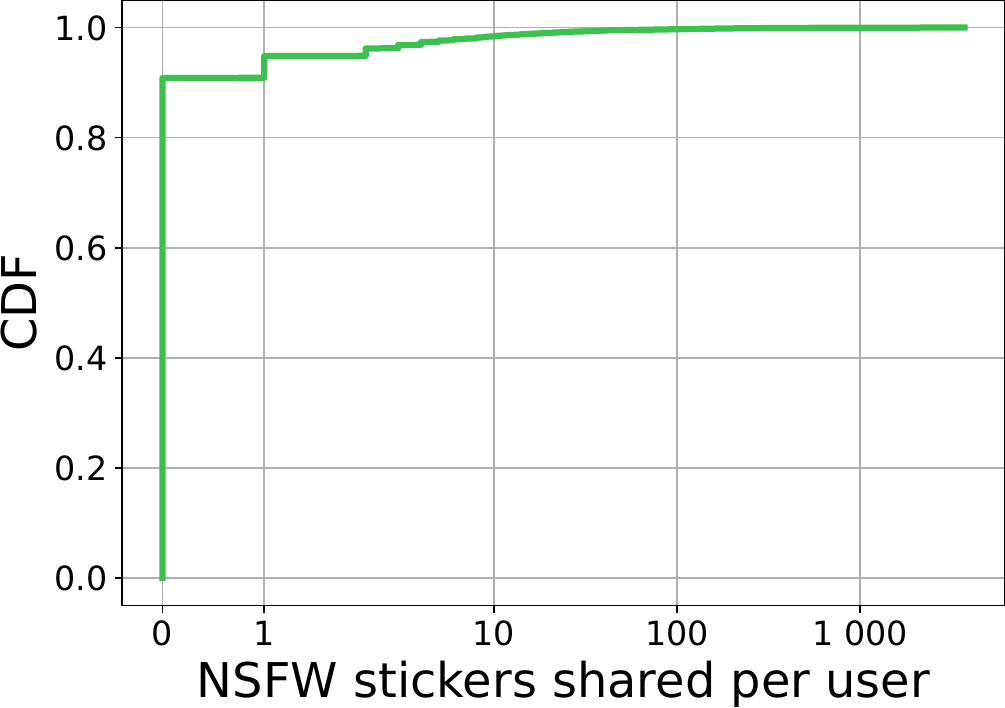}\label{fig:nsfw_stickers_users}}
    \caption{Distribution of NSFW stickers sent.}
    \label{fig:nsfw_stickers}
\end{figure}

In our data, we also found a considerable presence of stickers with sexually explicit content. To assess this, 
we apply the convolutional neural network model for Not Suitable for Work (NSFW)  proposed by Yahoo Inc.~\cite{nsfw_yahoo} to identify adult themed stickers. Images with a score greater than 80\% are labeled as explicit.
Figure \ref{fig:nsfw_stickers_images} presents the volume of NSFW stickers. Compared to images, stickers are five times more likely to depict explicit content (0.5\% of images and 2.3\% of stickers are NSFW). 
In total, we discovered 33,335 messages containing NSFW stickers, accounting for 5.5\% of all sticker messages. Notably, in Figure~\ref{fig:nsfw_stickers_users}, we see that approximately 10\% of users from the monitored groups shared a NSFW sticker, with some individuals posting hundreds of NSFW content. There is even a peculiar user who shared alone more than 3000 NSFW stickers. These results corroborate with the idea of abusive behavior on WhatsApp, especially considering the nature of the public political groups monitored are usually not NSFW (some of them even state rules in the description forbidding adult content on that chat).

\begin{table}[t]
    \centering
    \caption{``Not Safe'' stickers categories on WhatsApp.}
    \label{tab:abusive}
    \resizebox{\columnwidth}{!}{%
        \begin{tabular}{cll}
        \toprule    
        \multicolumn{1}{c}{ }
                                       & \multicolumn{1}{c}{\textbf{Unique Stickers}} & \multicolumn{1}{c}{\textbf{Number of Shares}} 
        \\ \midrule
        \multicolumn{1}{l}{Yahoo\_NSFW}    & 1,322  (2.3\%)                                  & 33,335 (5.05\%)                             \\ 
        \multicolumn{1}{l}{Adult}    & 1,423  (2.5\%)                                  & 30,828 (4.6\%)                             \\ 
        \multicolumn{1}{l}{Violence} & 416  (0.7\%)                                  & 4,583 (0.7\%)                              \\ 
        \multicolumn{1}{l}{Racy}     & 4,503 (7.9\%)                                 & 55,476 (8.4\%)                            \\ 
        \multicolumn{1}{l}{Medical}  & 735  (1.3\%)                                  & 14,300  (2.1\%)                            \\ 
        \multicolumn{1}{l}{Spoof}    & 17,834 (31.2\%)                               & 150,351 (22.7\%)                           
        \\ \bottomrule
        \end{tabular}
    }
\end{table}

Furthermore, we evaluate other categories of NSFW content through the Safe Search detection from Google's Vision API\footnote{\url{https://cloud.google.com/vision}}. It detects content within an image across five categories (adult, spoof, medical, violence, and racy) and returns the likelihood for each one of them.
According to the API, Adult content encompasses elements such as nudity, pornographic images, cartoons, and activities of a sexual nature. The medical category accounts for the likelihood that the content is associated with medical imagery (e.g. wounds, surgeries, blood). Violence offers an estimation of the probability that the image portrays a violent action.
Spoof is the likelihood that a modification was made to the image's canonical version to make it appear funny or offensive. Content deemed racy may include (but is not limited to) skimpy clothing and strategically covered nudity, lewd, provocative poses, and extreme close-ups of sensitive body areas. In Table~\ref{tab:abusive} we aggregated the results of Safe Search detection on our data collection of stickers considering the likely and very likely labels for each category.

These findings underscore the urgent need for greater moderation and enforcement of content guidelines within WhatsApp's sticker ecosystem. Unlike images, which users can choose not to automatically download, stickers are directly displayed to users in WhatsApp chats, making them susceptible to abuse by malicious actors. While stickers can enhance user engagement and communication~\cite{wang2016more}, their misuse brings significant risks, including promoting hate speech, perpetuating harmful stereotypes, and creating unsafe environments within online communities.

\section{Conclusion} \label{sec:conclusion}

Stickers are an emerging media format that has attained massive popularity on instant message platforms. 
With digital communication more present in our daily lives than ever, users demand more effective ways to express feelings and specific concepts within virtual chats. In this context, stickers represent a welcome and rapidly adopted media to extend the expressiveness of online conversation. Despite their widespread popularity, our understanding of sticker usage remains limited, primarily due to the closed and private nature of messaging app architectures. The opaque nature of these platforms hinders the comprehensive analysis of sticker and their implications for online discourse.

In this work, we analyze 659K stickers messages from WhatsApp political public groups comprising a total of 57,031 unique stickers shared during four months of the 2022 presidential Brazilian elections and offer insights into how stickers are employed in this specific domain. Our findings reveal some characteristics that suggest stickers are indeed a distinctive media format but share similarities with others: stickers are less frequently sent compared to other media types, such as images within groups. However, a single sticker usually presents higher total shares, which means they exhibit a higher rate of recurrence within conversations, similar to an emoji. 
Moreover, stickers are rarely forwarded, suggesting that users often save and keep their collection of preferred stickers for future use instead of using sharing tools of the system. This behavior highlights the collectible nature of stickers, in which individuals accumulate a personalized repertoire of stickers to deploy on various occasions like a meme.

Furthermore, our analysis investigates two distinct methodologies for categorizing stickers, each offering unique insights into their usage patterns. Initially, we employ a visual similarity approach, clustering stickers based on perceptual hashing -- pHash. This method reveals parallels of stickers with the usage of emojis and memes, showing how stickers can present minor variations while keeping the same essence of a base image template. However, we also uncover instances where visually similar stickers convey entirely different meanings, emphasizing the importance of considering other aspects when grouping images by perceptual hashes. Particularly in political groups, where misinformation dissemination is a concerning issue, it is important to avoid merging distinct ideas within the same category, especially when minor alterations in images could propagate misleading narratives.

Additionally, our second approach includes building a network of stickers based on their co-occurrence in public groups. This network analysis resulted in two prominent communities of stickers, mirroring the polarized partisan landscape of our political WhatsApp dataset. Subsequently, by labeling stickers according to their political alignment, we find a strong association between stickers and political affiliations. Notably, we observe instances where political stickers are shared across ideologically opposing groups, often serving as tools for provocation and inciting discord.

Moreover, our findings reveal the urgent need for better moderation mechanisms to curb the dissemination of offensive stickers and safeguard users from encountering inappropriate content on WhatsApp. Our analysis of the dataset reveals a notable prevalence of potentially offensive sticker messages, including content perpetuating homophobia, featuring hate symbols like the swastika, perpetrating racist stereotypes, and depicting derogatory illustrations of marginalized communities. We also discovered that stickers are five times more likely to depict explicit (NSFW) content compared to images; which represents 2.3\% of unique stickers and 5\% of all sticker messages containing sexually explicit content. Furthermore, we identified a substantial volume of sticker messages featuring violent (0.7\%), medical (2.1\%), and racy (8.4\%) content, which may be highly sensitive for certain audiences.  
Given the visual immediacy of stickers within chat interfaces, the unrestricted access to public groups, and the ease with which users can create and disseminate stickers from virtually any image, there are significant risks of offensive content spreading unchecked due to lack of moderation.

In conclusion, stickers have transformed communication on instant messaging platforms by offering users a nuanced way to express emotions and ideas. While they bring novelty to online chats, their impact on platform dynamics is largely unexplored. Despite WhatsApp's predominantly private and encrypted nature, our research provides valuable insights into how stickers are utilized within public political groups, bringing more transparency to the platform. We discover key differences that make a distinction between stickers and other more traditional formats. and analyzed the implications of the usage of stickers within political groups and their partisan nature. Finally, we identify a substantial abuse of stickers with offensive content on WhatsApp, which requires actions to bring more safety to users. 

As future work, we aim to investigate the use of stickers across different scenarios and subjects within WhatsApp. We also intend to further explore the prevalence and impact of hate stickers, seeking to identify patterns and strategies to mitigate their dissemination on the platform. Additionally, we plan to expand our analysis to other messaging platforms such as Telegram, broadening our understanding of sticker usage in different contexts and environments. By continuing to explore and address the complexities of stickers, we can contribute to building a safer and more informed online environment for users worldwide.

\footnotesize
\bibliography{aaai24}

\end{document}